\documentclass{mn2e}
\usepackage{graphicx}

\begin{document}
\title[Excitation of oscillation modes by tides in close
  binaries]{Excitation of oscillation modes by tides in close
  binaries: constraints on stellar and orbital parameters}
\author[B. Willems]
{B. Willems\thanks{E-mail: B.Willems@open.ac.uk} \\
Department of Physics and Astronomy, The Open University,
Walton Hall, Milton Keynes, MK7 6AA, UK}

\date{Accepted ... Received ...; in original form ...}

\pagerange{\pageref{firstpage}--\pageref{lastpage}} \pubyear{2003}

\maketitle

\label{firstpage}

\begin{abstract}
The parameter space favourable for the resonant excitation of free
oscillation modes by dynamic tides in close binary components is
explored using qualitative considerations to estimate the order of
magnitude of the tidal force and the frequency range covered by the
tidally induced oscillations. The investigation is valid for slowly
rotating stars with masses in the interval between 2 and
$20\,M_\odot$, and an evolutionary stage ranging from the beginning to
the end of the main sequence. Oscillation modes with eigenfrequencies
of the order of five times the inverse of the star's dynamical
time scale $\tau_{\rm dyn}$ - i.e. the lowest-order $p$-modes, the
$f$-mode, and the lowest-order $g^+$-modes - are found to be outside
the favourable parameter space since their resonant excitation
requires orbital eccentricities that are too high for the binary to
stay detached when the components pass through the periastron of their
relative orbit. Resonances between dynamic tides and $g^+$-modes with
frequencies of the order of half of the inverse of the star's 
dynamical time scale on the other hand are found to be favourable for
orbital periods up to $\sim 200\, \tau_{\rm dyn}$, provided that the
binary mass ratio $q$ is larger than 1/3 and the orbital eccentricity
$e$ is larger than $\sim 0.25$. This favourable range comes down 
to orbital periods of up to $5-12$ days in the case of 
$2-20\,M_\odot$ zero-age main-sequence binary components, and 
orbital periods of up to $21-70$ days in the case of terminal 
main-sequence binary components.
\end{abstract}


\section{Introduction}

In close binaries with non-synchronously rotating components, each
star experiences the time-dependent tidal force exerted by its
companion. The tidal force gives rise to forced nonradial oscillations
with frequencies that are determined by the orbital period and the
rotation rates of the component stars. The forcing frequencies may, at
any time, come close to the eigenfrequencies of the stars' free
oscillation modes because of secular changes in the orbital parameters
due to orbital and rotational angular momentum losses or because of
changes in the eigenfrequencies due to stellar evolution. Such
resonances lead to an enhanced tidal action and result in the
excitation of the oscillation mode involved in the resonance.

The resonant excitation of free oscillation modes by the tidal action
of a companion was first suggested in a seminal paper by Cowling
(1941). Since then, various authors have approached the problem from
both an analytical and a numerical point of view with the aim to
explain some of the evolutionary aspects and the observational
characteristics of binary stars (e.g. Zahn 1970, 1975; Savonije \&
Papaloizou 1983, 1984; Willems, Van Hoolst \& Smeyers 2003, and
references therein). The subject recently received a new impetus,
first from the discovery of giant planets in close orbits around
solar-type stars (see Willems et al. 1997, and references therein),
and subsequently from the increasingly accurate large-scale surveys of
pulsating stars in close binaries (e.g. Harmanec et al. 1997, Aerts et
al. 1998, Aerts et al. 2000). The potential of tides as an excitation
mechanism was furthermore found to be particularly promising for the
excitation of gravity modes in hot B subdwarfs by Fontaine et
al. (2003).  Existing and upcoming space missions such as WIRE, MOST,
MONS, COROT, and Eddington are likely to contribute even further to
the rising interest in tidally excited oscillation modes.

The occurrence of resonances and their effects on the evolution and
the observational properties of a binary depend on the period and the
eccentricity of the orbit, on the masses and the radii of the
component stars, and on the properties of the oscillation mode
involved in the resonance. The latter in turn depend sensitively on
the internal stellar structure of the tidally distorted star. The
multitude of these dependencies and the complexity of their
interaction with each other make it difficult to form a concise
intuitive picture on the possibility of exciting nonradial
oscillations by tides in close binaries. A systematic study
unravelling the role of the different stellar and orbital parameters
would therefore provide a valuable addition to the theory of tidally
induced oscillations as well as to observational campaigns dedicated
to the search of free or forced oscillations in binary stars.

In this paper, we set a first step in an attempt to understand the
parameter space of stellar and orbital parameters for which the
circumstances for the resonant excitation of free oscillation modes
are favourable. The primary aim is to clarify the qualitative
behaviour of the acting parameters and the interaction between them
without detailed calculations involving the numerical integration of
the systems of differential equations governing free and forced
oscillations in close binary components. The results of this study are
then to serve as a starting point for detailed numerical calculations
which will be presented in a follow-up investigation.

The plan of the paper is as follows. In Section~1, we present the
basic assumptions and the basic ingredients of the theory of forced
oscillations in close binaries. Some particular attention is paid to
the Fourier decomposition of the tide-generating potential, which
plays a crucial part in the investigation. In Section~2, we present a
detailed analysis of the Fourier coefficients as functions of the
orbital eccentricity. In Section~3, we examine the ratio of the tidal
force to the gravity at the star's equator. In Section~4, the results
of the preceding sections are used to constrain the range of forcing
frequencies that is favourable for the resonant excitation of free
oscillation modes by dynamic tides in close binaries. In the final
section, we summarise our results and present some concluding remarks.

\section{Basic concepts and assumptions}

Consider a close binary system of stars orbiting each other under the
influence of their mutual gravitational force. Let $P_{\rm orb}$ be
the orbital period, $a$ the semi-major axis, and $e$ the orbital
eccentricity. The first star, with mass $M_1$ and radius $R_1$, is
assumed to be a uniformly rotating main-sequence star with a rotation
axis perpendicular to the orbital plane. The magnitude of the
rotational angular velocity $\vec{\Omega}$ is assumed to be small in
comparison to the inverse of the star's dynamical time scale:
\begin{equation}
\Omega \ll {1 \over \tau_{\rm dyn}} \equiv \left( {{G M_1} \over R_1^3}
  \right)^{1/2},  \label{omega}
\end{equation}
where $G$ is the Newtonian constant of gravitation. The second star,
with mass $M_2$, is treated as a point mass.

Furthermore, let $\vec{r} = \left( r,\theta,\phi \right)$ be a system
of spherical coordinates with respect to an orthogonal frame of
reference that is corotating with the star, and let $\varepsilon_T\, W
\left( \vec{r},t \right)$ be the potential giving rise to the tidal
force. Here, $\varepsilon_T$ is a small dimensionless parameter
which is proportional to the ratio of the tidal force to the
gravity at the star's equator and is defined as
\begin{equation}
\varepsilon_T = \left( {R_1 \over a} \right)^3 {M_2 \over M_1}
  \label{epsT}
\end{equation}
(e.g. Tassoul 1987).

The tide-generating potential can be expanded in terms of unnormalised
spherical harmonics $Y_\ell^m(\theta,\phi)$ and in Fourier series in
terms of multiples of the companion's mean motion $n=2\,\pi/P_{\rm
orb}$ as
\begin{eqnarray}
\lefteqn{\varepsilon_T\, W \left( \vec{r},t \right) = -
  \varepsilon_T\, {{G M_1} \over R_1}\,
  \sum_{\ell=2}^4 \sum_{m=-\ell}^\ell \sum_{k=-\infty}^\infty}
  \nonumber \\
& & c_{\ell,m,k}\, \left( {r \over R_1} \right)^\ell
  Y_\ell^m (\theta,\phi)\, \exp \left[ {\rm i}
  \left( \sigma_T\, t - k\, n\, \tau \right) \right],
  \label{pot}
\end{eqnarray}
where $\sigma_T = k\, n + m\, \Omega$ is a forcing angular frequency
with respect to the corotating frame of reference and $\tau$ is a time
of periastron passage. The Fourier coefficients $c_{\ell,m,k}$ are
determined by
\begin{eqnarray}
\lefteqn{c_{\ell,m,k} = \displaystyle
  {{(\ell-|m|)!} \over {(\ell+|m|)!}}\, P_\ell^{|m|}(0)
  \left({R_1\over a}\right)^{\ell-2}
  {1\over {\left({1 - e^2}\right)^{\ell - 1/2}}} } \nonumber \\
 & & {1\over \pi} {\int_0^\pi (1 + e\, \cos v)^{\ell-1}\,
  \cos (k\, M + m\, v)\, dv}, \label{pot:2}
\end{eqnarray}
where $P_\ell^{|m|}(x)$ is an associated Legendre polynomial of the
first kind, and $M$ and $v$ are respectively the mean and the true
anomaly of the companion in its relative orbit (e.g. Polfliet \&
Smeyers 1990; Smeyers, Willems \& Van Hoolst 1998). The Fourier
coefficients obey the property of symmetry $c_{\ell,-m,-k} =
c_{\ell,m,k}$ and are different from zero only for even values of
$\ell+|m|$. In the particular case of a binary with a circular orbit
they are different from zero only for $k=-m$.

It follows that, for each degree $\ell$ of the spherical harmonics
$Y_\ell^m (\theta,\phi)$, the tide-generating potential generates an
infinite number of partial dynamic tides with forcing angular
frequencies $\sigma_T \ne 0$. When one of these forcing frequencies is
close to the eigenfrequency of a free oscillation mode, the tidal
action exerted by the companion is enhanced and the oscillation mode
is resonantly excited. The occurrence of resonances between partial
dynamic tides and free oscillation modes is particularly relevant for
the excitation of free oscillation modes $g^+$ since, for binaries
with short orbital periods, their eigenfrequencies may be in the range
of the forcing frequencies induced by the companion.

Since the terms associated with the third- and fourth-degree spherical
harmonics in the expansion of the tide-generating potential contain an
additional small factor $R_1/a$ or $\left( R_1/a \right)^2$ with
respect to the second-degree terms, the latter ones are usually
dominant. For the remainder of the paper, we therefore restrict
ourselves to the terms associated with $\ell=2$. The only non-zero
Fourier coefficients in Expansion~(\ref{pot}) of the tide-generating
potential are then the coefficients $c_{2,-2,k}$, $c_{2,0,k}$, and
$c_{2,2,k}$. These coefficients are independent of the ratio $R_1/a$
so that they are solely determined by the orbital eccentricity
$e$. Since they also obey the symmetry property $c_{2,-m,-k} =
c_{2,m,k}$, it is furthermore sufficient to consider positive values
of $k$ only.

The effects of resonant dynamic tides on the evolution and the
observational characteristics of a binary depend on the strength of
the tide and on the properties of the oscillation mode involved in the
resonance (e.g. Smeyers et al. 1998, Willems et al. 2003). The
strength of the tide is determined by the mass ratio $q=M_2/M_1$ and
the orbital separation $a$ via the dimensionless parameter
$\varepsilon_T$, and by the orbital eccentricity $e$ via the Fourier
coefficients $c_{\ell,m,k}$. The properties of the oscillation mode
contribute to the effects of the resonance via the so-called
overlap-integral which is proportional to the ratio of the work done
by the tidal force to the mode's kinetic energy (for a precise
definition see, for example, Press \& Teukolsky 1977, Kumar, Ao \&
Quataert 1995, Smeyers et al. 1998). For main-sequence stars, the
behaviour of the overlap-integral is such that the coupling between
the oscillation mode and the tidal force generally weakens with
increasing radial order of the mode.

In the following sections, we focus on the role of the stellar and
orbital parameters affecting the strength and the frequency of the
tides. The role of the internal structure of the star and the
properties of the oscillation modes will be the subject of a
forthcoming investigation.

\section{The orbital eccentricity}
\label{ecc}

The orbital eccentricity contributes to the determination of the
strength of the partial dynamic tides through the Fourier coefficients
$c_{2,m,k}$. For a given eccentricity, the coefficients generally
decrease with increasing values of $k$, but the decrease is slower for
higher orbital eccentricities. In addition, for a given value of $k$,
the coefficients $c_{2,m,k}$ generally increase with increasing values
of $e$. Hence, there is a finite number of non-trivially
contributing terms in the expansion of the tide-generating potential
which increases with increasing orbital eccentricities.

\begin{figure*}
\resizebox{8.8cm}{!}{\includegraphics{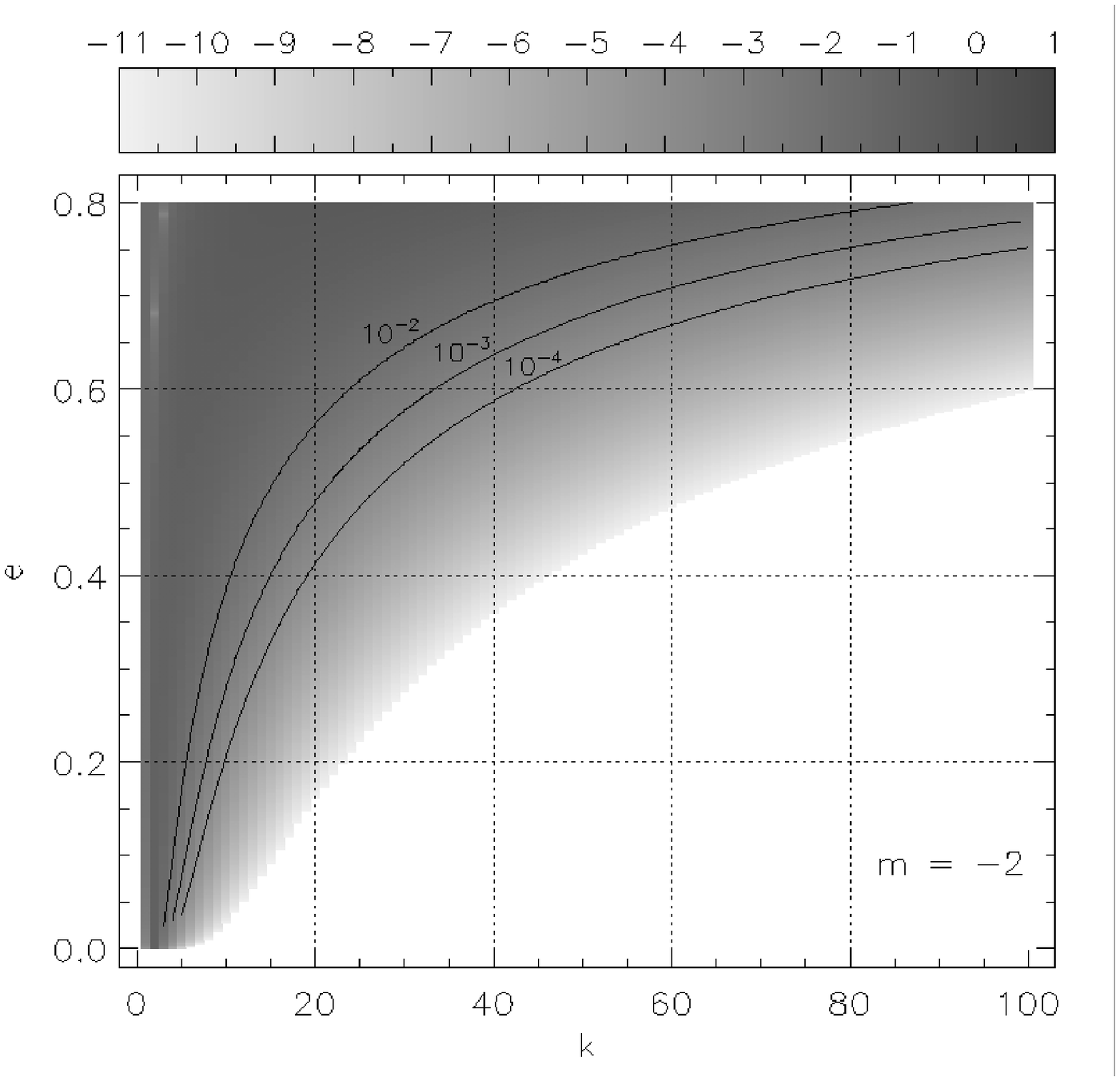}}
\resizebox{8.8cm}{!}{\includegraphics{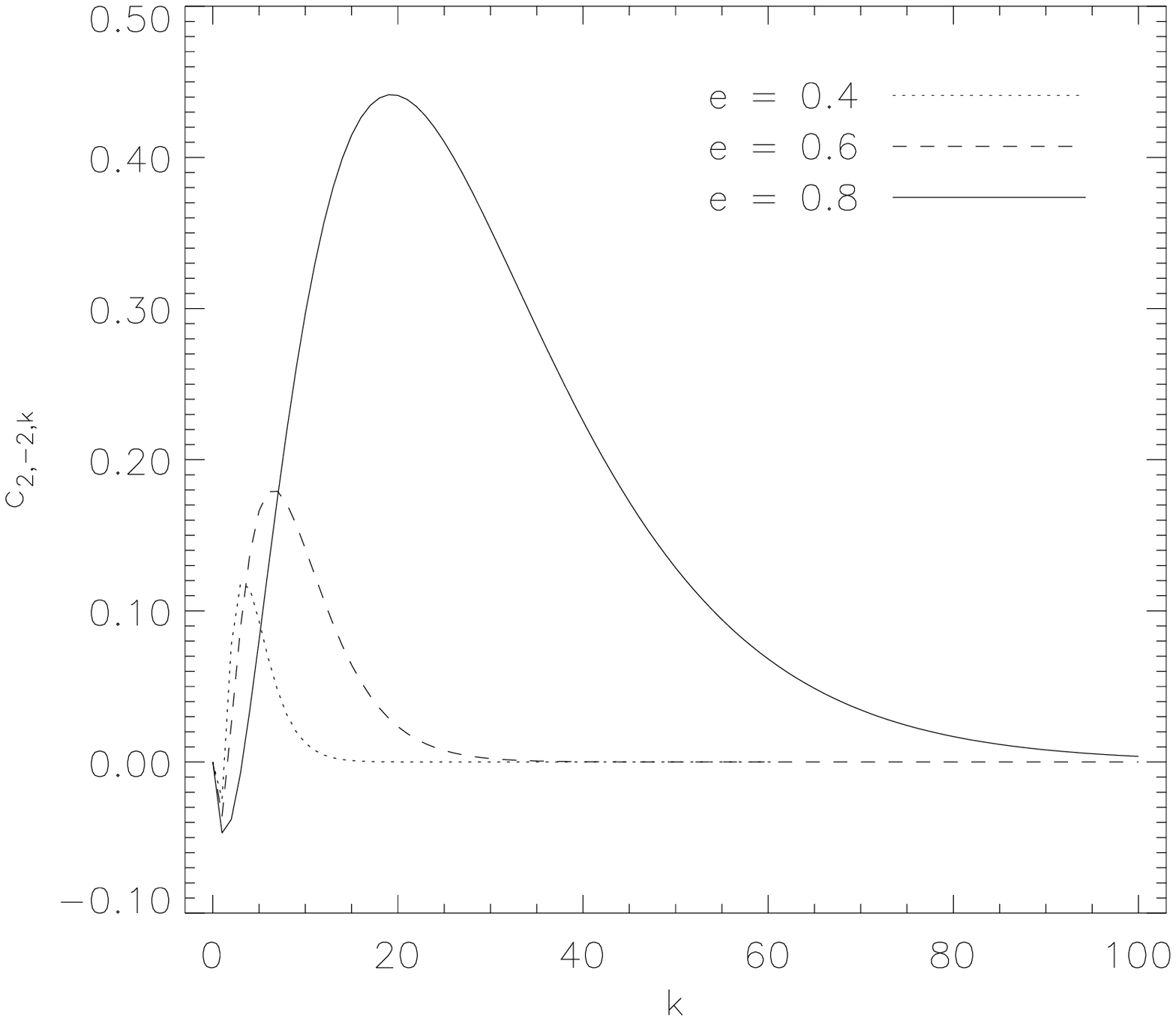}}
\caption{{\it Left:} The logarithm of the absolute value of the
  coefficients $c_{2,-2,k}(e)$ as a two-dimensional function of the
  Fourier index $k$
  and the orbital eccentricity~$e$. The solid lines indicate the
  values of $k$ and $e$ for which $|c_{2,-2,k}(e)| \approx 10^{-2}$,
  $10^{-3}$, $10^{-4}$. {\it Right:} Cross-section of the variation of
  the coefficients $c_{2,-2,k}(e)$ for the orbital eccentricities
  $e=0.4,0.6, 0.8$.}
\label{c-2}
\end{figure*}

\begin{figure*}
\resizebox{8.8cm}{!}{\includegraphics{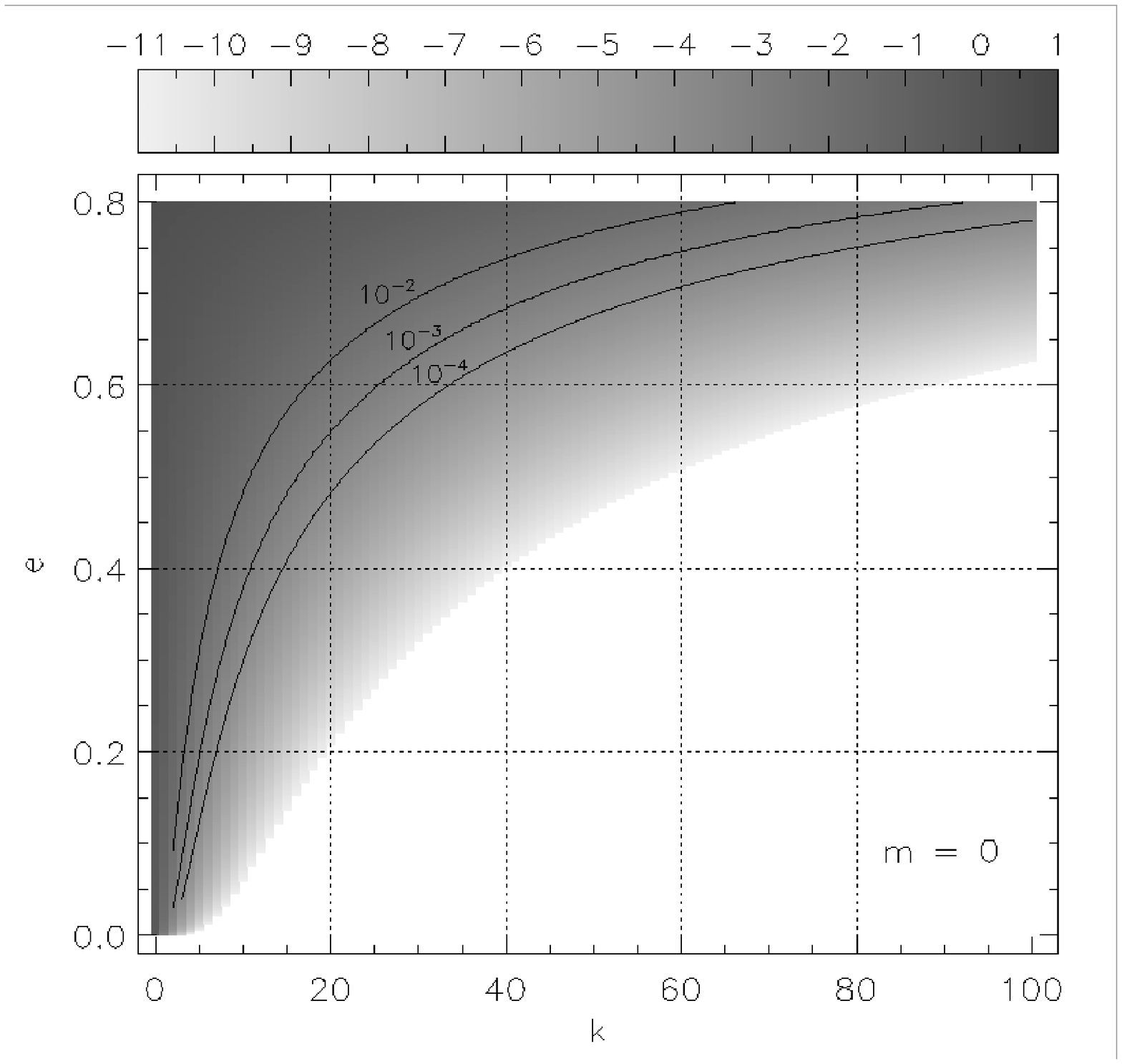}}
\resizebox{8.8cm}{!}{\includegraphics{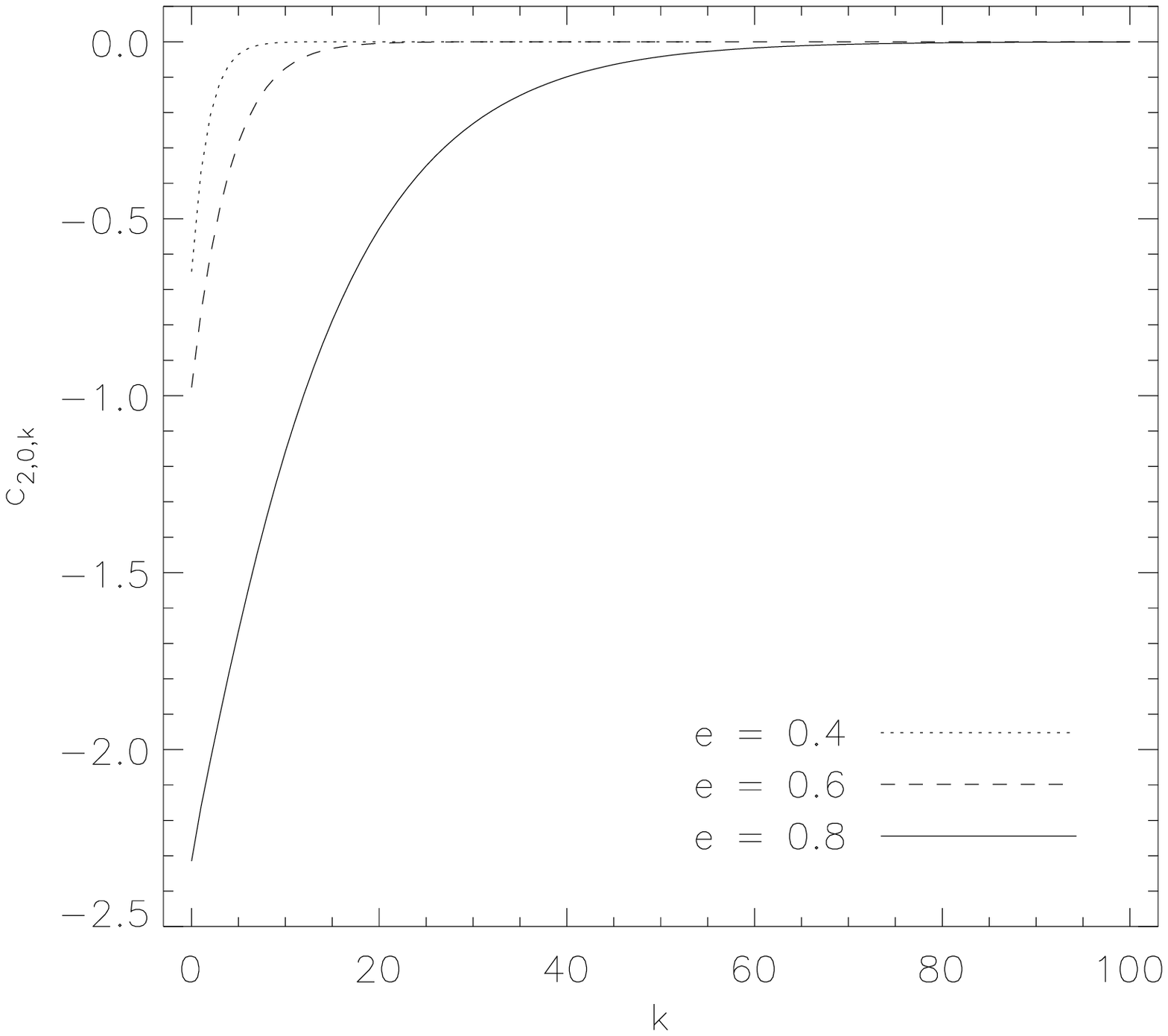}}
\caption{As Fig.~\ref{c-2}, but for the coefficients
  $c_{2,0,k}(e)$.}
\label{c0}
\end{figure*}

\begin{figure*}
\resizebox{8.8cm}{!}{\includegraphics{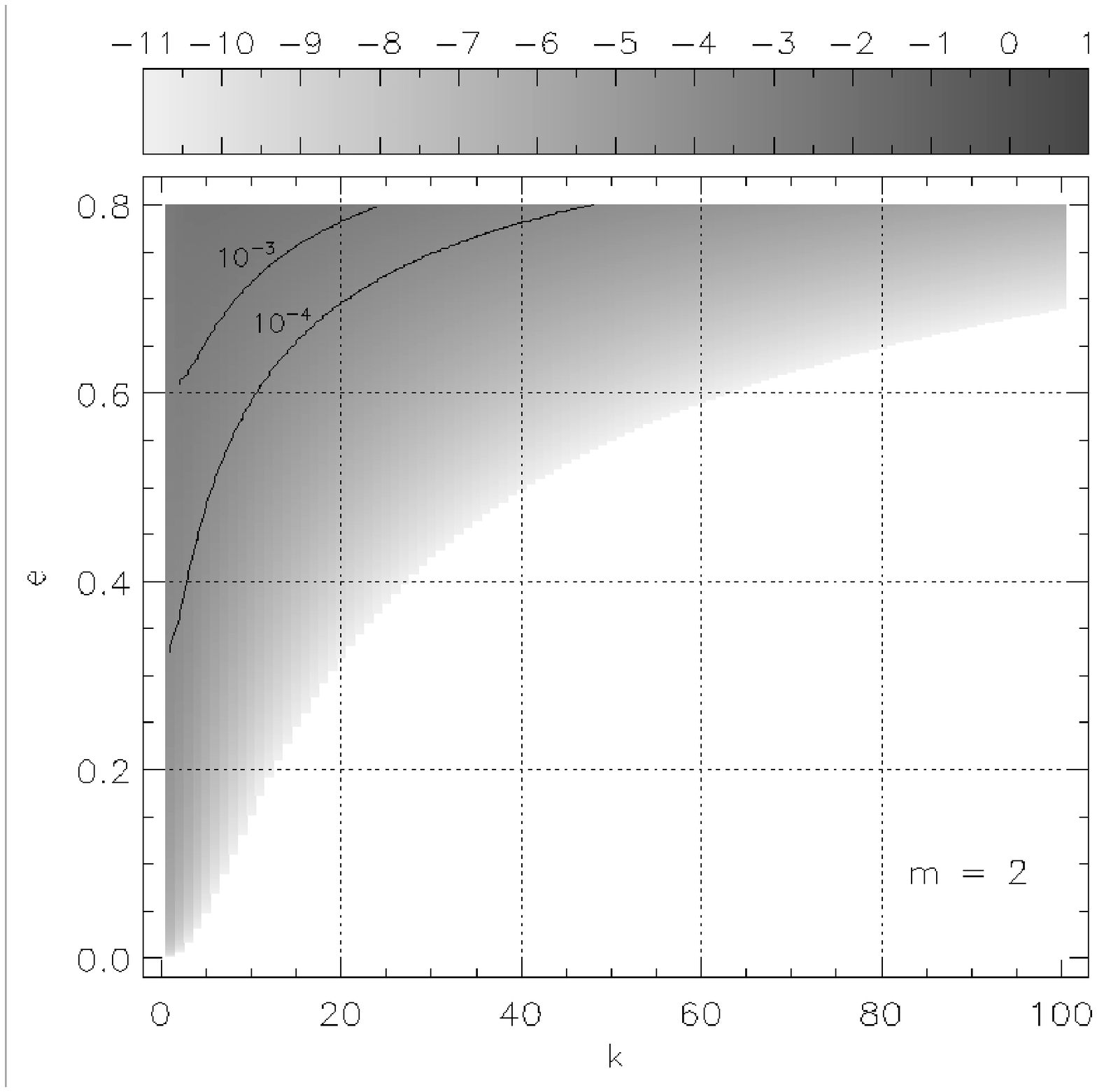}}
\resizebox{8.8cm}{!}{\includegraphics{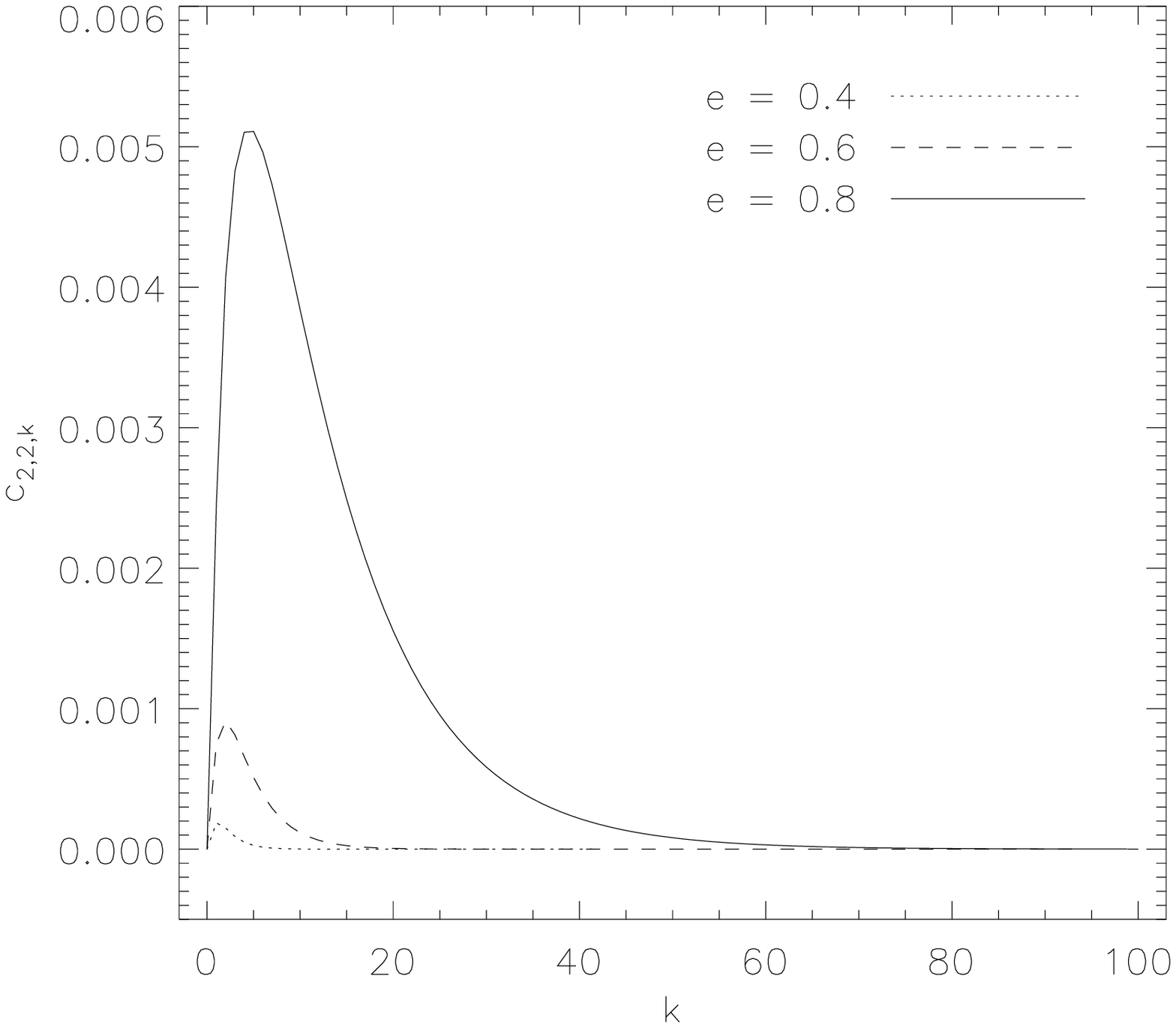}}
\caption{As Fig.~\ref{c-2}, but for the coefficients
  $c_{2,2,k}(e)$.}
\label{c2}
\end{figure*}

In order to estimate the value of $k$ beyond which the contributions
to the expansion of the tide-generating potential become negligible as
a function of the orbital eccentricity, the logarithms of the absolute
values of the non-zero coefficients $c_{2,m,k}$ are displayed in the
left-hand panels of Figs.~\ref{c-2}--\ref{c2} as a two-dimensional
function of $k$ and $e$. For clarity, lines for which $|c_{2,m,k}(e)|
\approx 10^{-2}, 10^{-3}, 10^{-4}$ are added to the left-hand panels
of the figures and one-dimensional variations of the coefficients
$c_{2,m,k}$ as a function of $k$ are shown in the right-hand panels in
the case of the orbital eccentricities $e=0.4, 0.6$, and $0.8$. The
smallest Fourier coefficients considered have absolute values of the
order of $10^{-11}$.

As mentioned in the beginning of this section, the general tendency of
the coefficients $c_{2,m,k}$ is to decrease rapidly with increasing
values of $k$. The behaviour is slightly more complex for the
coefficients $c_{2,-2,k}$ for which, depending on the value of the
orbital eccentricity, a local minimum and a local maximum may occur
prior to the rapid decrease with increasing values of
$k$\footnote{Note that the local maximum occurs at values of $k$ close
  to those for which $k\,n \approx \Omega_{\rm P}$, with $\Omega_{\rm
  P}= n \left[ \left(1+e\right)/\left(1-e\right)^3 \right]^{1/2}$ the
 orbital angular velocity at the periastron of the relative orbit.}. This
is particularly clear in the cross sections shown in the right-hand
panel of Fig.~\ref{c-2}. A similar behaviour is observed for the
coefficients $c_{2,2,k}$, albeit somewhat less pronounced. For a given
value of the orbital eccentricity $e$ and sufficiently large values of
the Fourier index $k$, the largest contributions to the expansion of
the tide-generating potential usually stem from the terms associated
with the coefficients $c_{2,-2,k}$, while the contributions of the
terms associated with the coefficients $c_{2,2,k}$ are usually
negligible. For large orbital eccentricities and small values of $k$,
the dominant contributions generally stem from the terms associated
with the coefficients $c_{2,0,k}$.

Although the value of $k$ beyond which the contributions to the
expansion of the tide-generating potential are considered to become
negligible depends on the desired accuracy of the Fourier
decomposition, a reasonable estimate for a cut-off value can be
obtained by observing the rapid decrease of the absolute value of the
Fourier coefficients from values of the order of $10^{-2}$ to values
of the order of $10^{-4}$. If we focus on the coefficients associated
with $m=-2$, the maximum $k$-value that needs to be considered
increases from $k \approx 5-10$ in the case of the orbital
eccentricity $e=0.2$ to $k \approx 25-45$ in the case of the orbital
eccentricity $e=0.6$. In the latter case there are thus up to 5 times
more forcing frequencies induced in the star which may be resonant
with the star's free oscillation modes $g^+$.

\section{The parameter $\varepsilon_T$}
\label{et}

The small dimensionless factor $\varepsilon_T$ in the definition of
the tide-generating potential can be used as a first estimate for the
order of magnitude of the tidal force exerted on a binary component by
its companion. In particular, the largest term in
Expansion~(\ref{pot}) of the tide-generating potential, evaluated at
the star's surface, is of the order of
\begin{equation}
\varepsilon_T\, {{G M_1} \over R_1}\, \max_{m,k} |c_{2,m,k}(e)|,
  \label{ord1}
\end{equation}
where the maximum is taken over all admissible values of $m$ and
$k$. For orbital eccentricities less than or equal to 0.8 the maximum
takes values between 0.5 and 2.5 [see Eq.~(\ref{pot:2}) and
Figs.~\ref{c-2}--\ref{c2}]. The eccentricity therefore affects the order
of magnitude by less than a factor of 2.5.

Definition~(\ref{epsT}) of the small parameter $\varepsilon_T$ can
be rewritten by means of Kepler's third law as
\begin{equation}
\varepsilon_T = {{4\, \pi^2} \over {P_{\rm orb}^2}}\,
  {{R_1^3} \over {G\,M_1}}\, {q \over {1+q}}.  \label{epsT2}
\end{equation}
The parameter is thus proportional to the square of the star's dynamic
time scale and inversely proportional to the square of the orbital
period. In addition, it depends on the mass ratio $q$ by the factor
$q/(1+q)$ which is always smaller than 1. In the case of a binary
consisting of two $10\,M_\odot$ zero-age main-sequence stars with an
orbital period of 6\,days, the parameter $\varepsilon_T$ is of the order
of $10^{-3}$.

For detached binaries, an upper limit for $\varepsilon_T$ can be
derived from the requirement that the radius of the star must be
smaller than the volume equivalent radius of its Roche lobe. The
latter may be approximated by means of Eggleton's (1983) fitting
formula
\begin{equation}
{R_{{\rm L},1} \over a} = {{0.49\, q^{-2/3}} \over {0.6\, q^{-2/3}
  + \ln \left( 1 + q^{-1/3} \right)}},  \label{RL}
\end{equation}
where $q= M_2/M_1$. The relative error of the approximation is smaller
than 2\% for $0<q<\infty$. By means of Definition~(\ref{epsT}) and
Eq.~(\ref{RL}), one then derives the inequality
\begin{equation}
\varepsilon_T < {{0.12} \over {q \left[ 0.6\, q^{-2/3}
  + \ln \left( 1 + q^{-1/3} \right) \right]^3}}.  \label{epsT3}
\end{equation}
Taking into account the 2\% uncertainty in the fitting formula given
by Eq.~(\ref{RL}), it follows that $0 < \varepsilon_T < 0.13$.

\begin{figure}
\resizebox{\hsize}{!}{\includegraphics{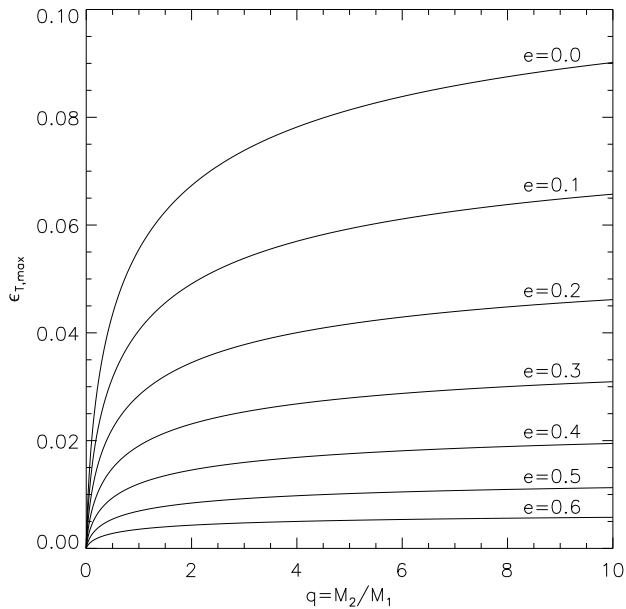}}
\caption{The maximum values of the small parameter $\varepsilon_T$ for
  which a binary with an orbital eccentricity $e$ and a mass ratio $q$
  is still detached.}
\label{epsTmax}
\end{figure}

Equation~(\ref{RL}) for the Roche-lobe radius of a binary component is
strictly speaking only valid for binaries with circular orbits. In the
case of a binary with an eccentric orbit, the radius of the Roche lobe
varies periodically in time, with the smallest value being reached
when the companion passes through the periastron of its relative
orbit. An instantaneous estimate for the Roche-lobe radius at a time of
periastron passage is obtained by replacing the semi-major axis $a$ in
Eq.~(\ref{RL}) with the periastron distance $r_p=a(1-e)$. The
requirement that the star fits within its Roche-lobe when the
companion is located in the periastron of its relative orbit then
yields
\begin{equation}
\varepsilon_T < {{0.12\, (1-e)^3} \over {q \left[ 0.6\, q^{-2/3}
  + \ln \left( 1 + q^{-1/3} \right) \right]^3}}.  \label{epsT5}
\end{equation}
The variation of the right-hand member of this inequality as a
function of the mass ratio $q$ is illustrated in Fig.~\ref{epsTmax}
for orbital eccentricities ranging from 0.0 to 0.6. It follows that,
for a given value of $e$, the parameter $\varepsilon_T$ takes values
in the range $0 < \varepsilon_T < 0.13\, (1-e)^3$. The maximum value
of $\varepsilon_T$ that can be reached by detached binaries decreases
by approximately one order of magnitude when $e$ increases from 0 to
0.5, and by approximately two orders of magnitude when $e$ increases
from 0 to 0.8.

\begin{figure}
\resizebox{8.8cm}{!}{\includegraphics{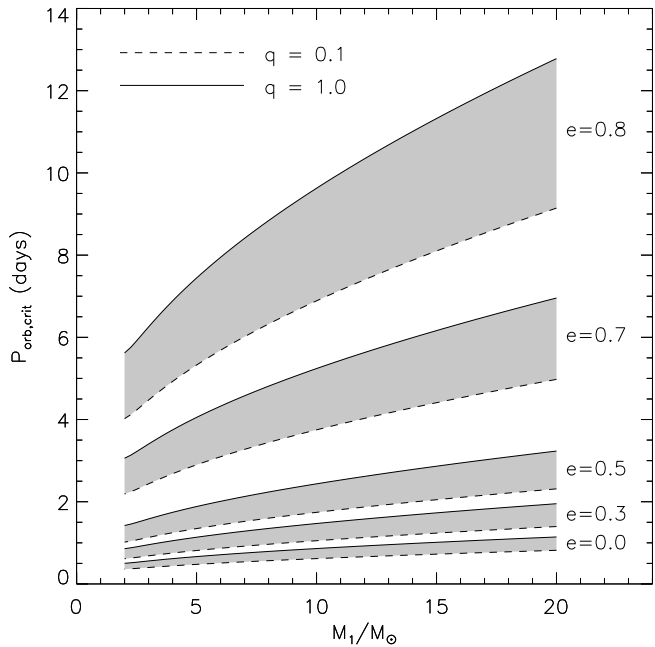}}
\caption{The critical orbital periods separating detached from
  semi-detached systems for $2-20\,M_\odot$ ZAMS stars and for binary
  mass ratios $q=0.1$ (dashed lines) and $q=1$ (solid lines). The
  grey-shaded region in between the lines associated with the two
  different mass ratios connects the lines associated with the same
  orbital eccentricity.}
\label{epsTmax_zams}
\end{figure}

\begin{figure}
\resizebox{8.8cm}{!}{\includegraphics{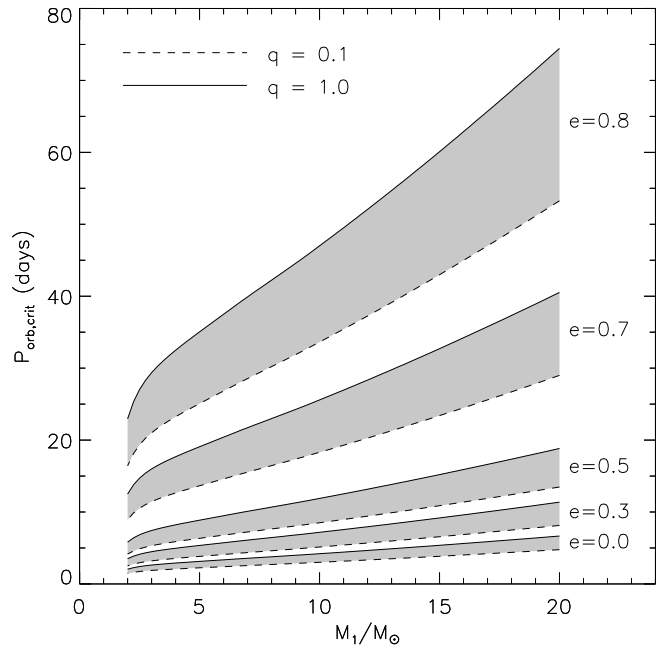}}
\caption{As Fig.~\ref{epsTmax_zams}, but for TMS stars.}
\label{epsTmax_msmax}
\end{figure}

Equations~(\ref{epsT2}) and~(\ref{epsT5}) can furthermore be combined
to derive the shortest orbital period for which a binary is still
detached when the components are located at the periastron of their
relative orbit. The resulting critical orbital periods separating
detached from semi-detached systems are displayed in
Figs.~\ref{epsTmax_zams} and~\ref{epsTmax_msmax} for $2-20\,M_\odot$
main-sequence stars at the beginning (ZAMS) and at the end (TMS) of
the main sequence, respectively.  The dashed lines represent the
critical orbital periods in the case of the binary mass ratio $q=0.1$,
while the solid lines represent the critical orbital periods in the
case of the binary mass ratio $q=1$. The grey-shaded region in between
the lines associated with the two mass ratios only serves as a visual
aid to identify the lines associated with the same orbital
eccentricity. We furthermore note that, for brevity, we here neglect
the short phase of core contraction that takes place between the
exhaustion of hydrogen in the core and the start of the Hertzsprung
gap, and we refer to the stage where hydrogen is exhausted in the core
as the end of the main sequence. The ZAMS and TMS stellar radii
appearing in Eq.~(\ref{epsT2}) are determined by means of the analytic
approximation formulae derived by Hurley, Pols \& Tout (2000).

Since the critical orbital periods separating detached from
semi-detached systems are proportional to the star's dynamical time
scale [see Eqs.~(\ref{epsT2}) and~(\ref{epsT5})], they increase with
increasing mass of the star. For a given orbital eccentricity and a
given mass ratio, the same dependency yields longer orbital periods
for stars on the TMS than for stars on the ZAMS. In addition, the
critical orbital periods increase with increasing values of the mass
ratio $q$ because the inner Lagrangian point $L_1$ moves closer to the
star with mass $M_1$. In the next section, we will use the behaviour
of the critical periods to constrain the range of forcing frequencies
induced by the tidal force exerted by the companion.

\section{The forcing frequencies}
\label{ff}

In order for a free oscillation mode $g^+$ to be resonantly excited by
a dynamic tide, the forcing frequency of the tide must be close to the
mode's eigenfrequency. Since the forcing frequencies depend on the
orbital period and the rotational angular velocity of the tidally
distorted star, the requirement that the forcing frequencies be in the
range of the eigenfrequencies of the lower-order $g^+$-modes imposes
constraints on the parameter space available for the occurrence
resonances.

\begin{figure}
\resizebox{\hsize}{!}{\includegraphics{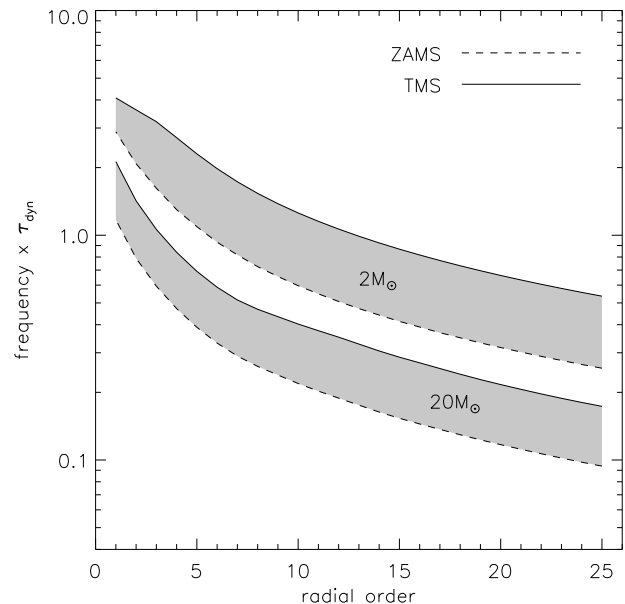}}
\caption{The frequency range covered by the eigenfrequencies of the
  second-degree $g^+$-modes of radial orders $N=1,2,\ldots,25$ in
  main-sequence stars with a mass between $2\,M_\odot$ and
  $20\,M_\odot$ and an evolutionary state between the ZAMS (dashed
  lines) and the TMS (solid lines). The frequencies are expressed in
  units of the inverse of the star's dynamical time scale.}
\label{freqs}
\end{figure}

The eigenfrequencies of the lower-order second-degree $g^+$-modes in
$2\,M_\odot$ and $20\,M_\odot$ main-sequence stars at the beginning
and at the end of the main sequence are shown in Fig.~\ref{freqs}, in
units of the inverse of the star's dynamical time scale.  The stellar
models consist of a convective core and a radiative envelope and have
a ZAMS chemical composition $(X,Z)=(0.7,0.02)$.  Since, for main-sequence
stars, the influence of nonadiabatic effects on the oscillation
frequencies is known to be small (e.g. Unno et al. 1989), the
eigenfrequencies are determined in the isentropic approximation.

The frequencies of the lower-order $g^+$-modes typically range from
$0.1\, \tau_{\rm dyn}^{-1}$ to $10\, \tau_{\rm dyn}^{-1}$, so that the
condition that the forcing frequencies be in the range of the
eigenfrequencies can be expressed as
\begin{equation}
k\,n + m\,\Omega \approx f\,\tau_{\rm dyn}^{-1}  \label{rng1}
\end{equation}
with $0.1 \la f \la 10$. For rotational angular velocities $\Omega \ll
\tau_{\rm dyn}^{-1}$, the $k$th harmonic in the expansion of the
tide-generating potential  may thus give rise to resonances with
lower-order $g^+$-modes for orbital periods
\begin{equation}
P_{\rm orb} \approx {{2\, \pi\, k} \over f}\, \tau_{\rm dyn}.
  \label{rng2}
\end{equation}

\begin{figure*}
\resizebox{\hsize}{!}{\includegraphics{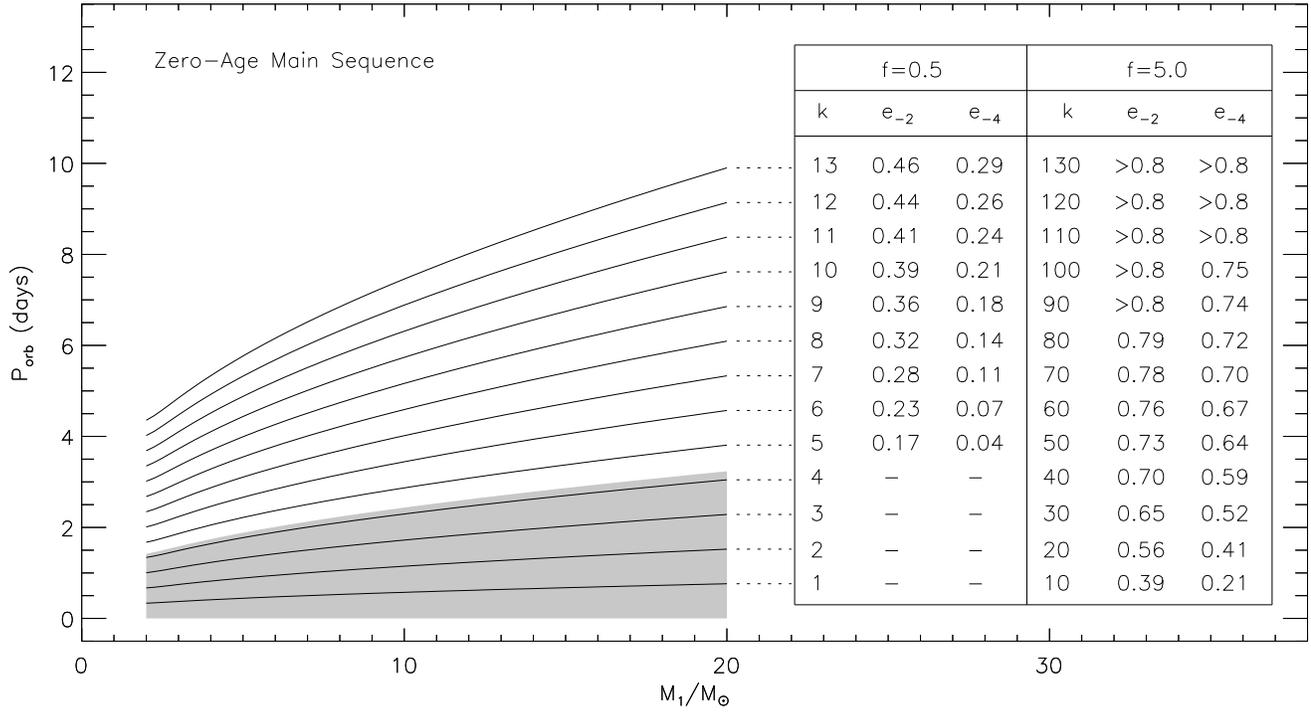}}
\caption{The orbital period range that may give rise to resonances
  with lower-order $g^+$-modes in $2-20\,M_\odot$ ZAMS stars. The
  solid lines correspond to the orbital periods obtained from
  Eq.~(\ref{rng2}) by setting $f=0.5$ and $k=1, 2, 3, \ldots, 13$; or
  $f=5.0$ and $k=10, 20, 30, \ldots, 130$. For each value of $k$, the
  table on the right-hand side of the figure lists the minimum values
  of the orbital eccentricity $e_{-2}$ and $e_{-4}$ for which
  $|c_{2,-2,k}(e)| \ga 10^{-2}$ and $|c_{2,-2,k}(e)| \ga 10^{-4}$,
  respectively. The grey-shaded background indicates the orbital
  periods for which a binary with a mass ratio $q=1$ and an orbital
  eccentricity $e=0.5$ becomes semi-detached when the stars pass
  through the periastron of their relative orbit.}
\label{M1Pzams}
\end{figure*}

\begin{figure*}
\resizebox{\hsize}{!}{\includegraphics{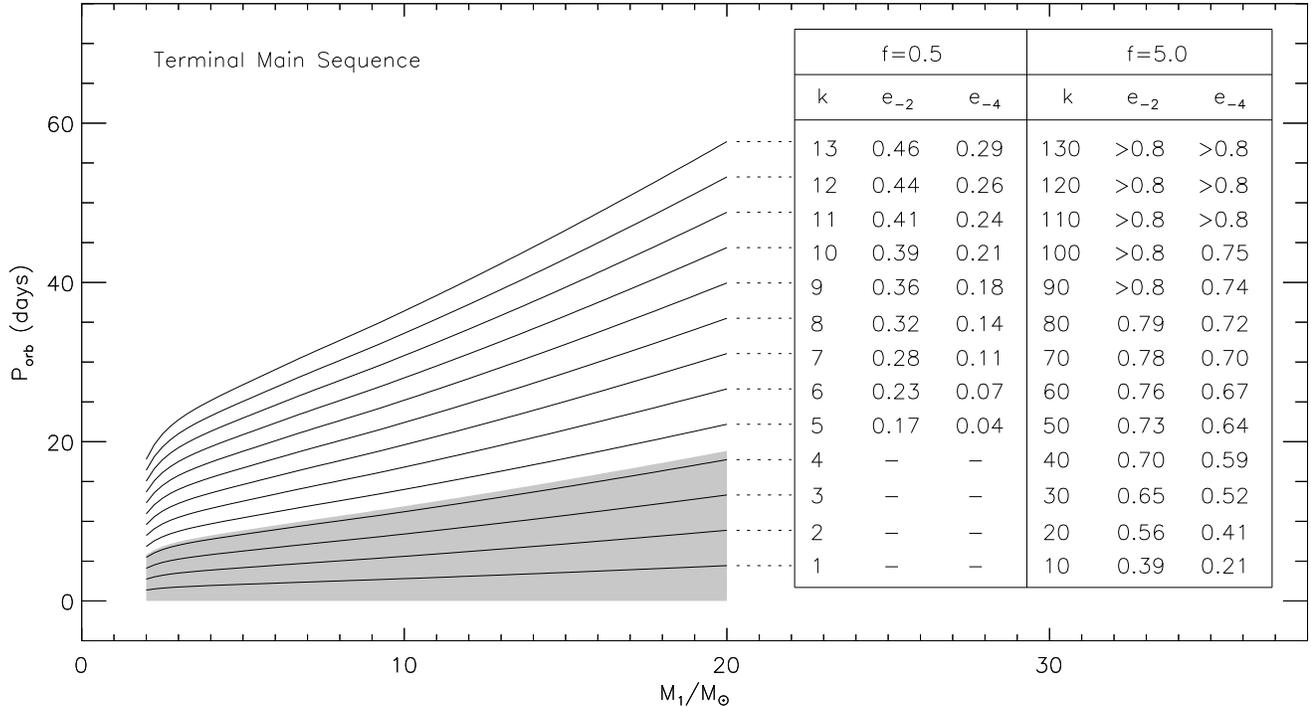}}
\caption{As Fig.~\ref{M1Pzams}, but for TMS stars.}
\label{M1Pmsmax}
\end{figure*}

In order to constrain the forcing frequencies leading to significant
resonances between dynamic tides and free oscillation modes, we
consider Eq.~(\ref{rng2}) in the cases where $f=0.5$ and $f=5.0$. The
former choice represents the bulk of the lower-order $g^+$-modes
displayed in Fig.~\ref{freqs}, while the latter choice represents an
upper limit on the frequency range of the $g^+$-modes. For
presentation purposes, the two values of $f$ are conveniently chosen
to differ by a factor of 10. The choice is convenient because, for a
given stellar model, the orbital periods associated with $k$ and $f$
are equal to those associated with $10\,k$ and $10\,f$. The variations
of the resulting orbital periods as a function of the stellar mass are
displayed by the solid lines in Figs.~\ref{M1Pzams}
and~\ref{M1Pmsmax}, for stars at the beginning and at the end of the
main sequence, respectively. The value of $k$ corresponding to each
line is indicated in the tables on the right-hand side of the
figures. It follows that, for ZAMS stars in binaries with orbital
periods of the order of a few days, the lower-order harmonics in
Expansion~(\ref{pot}) of the tide-generating potential give rise to
forcing frequencies of the order of $\sim 0.5\, \tau_{\rm dyn}^{-1}$,
while higher-order harmonics give rise to forcing frequencies of the
order of $\sim 5.0\, \tau_{\rm dyn}^{-1}$. For stars on the TMS, the
corresponding orbital period range increases to a few tens of days due
to the longer dynamical time scales associated with these stars.

Since the small parameter $\varepsilon_T$ appearing in the
definition of the tide-generating potential is inversely proportional
to the square of the orbital period, the tidal force decreases
rapidly with increasing orbital separations. From Eqs.~(\ref{epsT2})
and~(\ref{rng2}), one derives that the values of $\varepsilon_T$
associated with the orbital periods represented by the solid lines in
Figs.~\ref{M1Pzams} and~\ref{M1Pmsmax} are given by
\begin{equation}
\varepsilon_T = {f^2 \over k^2}\, {q \over {1+q}}.  \label{epsT6}
\end{equation}
For binaries with mass ratio $q=1$, the parameter $\varepsilon_T$
takes the value $10^{-2}$ when $f=0.5$ and $k \approx 3-4$ (or $f=5.0$
and $k \approx 35-36$) and decreases to $10^{-3}$ when $f=0.5$ and $k
\approx 11-12$ (or $f=5.0$ and $k \approx 111-112$). Smaller mass
ratios yield smaller values of $k$.

As discussed in Section~\ref{ecc}, the rapid decrease of the Fourier
coefficients $c_{2,m,k}$ in the expansion of the tide-generating
potential puts an upper limit on the maximum value of $k$ that needs
to be considered for a given value of the orbital eccentricity
$e$. Vice versa this implies that for each value of $k$ there is a
minimum orbital eccentricity for which the term associated with the
Fourier coefficient $c_{2,m,k}$ provides a non-negligible contribution
to the expansion of the tide-generating potential. We determined this
minimum value for each value of $k$ listed in the tables on the
right-hand side of Figs.~\ref{M1Pzams} and~\ref{M1Pmsmax}. To this
end, we restricted ourselves to the terms associated with the
azimuthal number $m=-2$ since they generally provide the largest
contributions. We furthermore adopted two different thresholds to
separate contributing from non-contributing terms:
$|c_{2,-2,k}(e)| \approx 10^{-2}$ and $|c_{2,-2,k}(e)| \approx
10^{-4}$ (cfr.\ the solid lines in Fig.~\ref{c-2}). The resulting
values of $e$ are listed in the columns labelled $e_{-2}$ and
$e_{-4}$, respectively. The table shows, for example, that the minimum
orbital eccentricity required for the absolute value of the Fourier
coefficient $c_{2,-2,5}$ to be larger than or equal to $10^{-2}$, and
thus for the associated term in the expansion of the tide-generating
potential to be non-negligible, is $e=0.17$. When the absolute value
of the Fourier coefficient $c_{2,-2,5}$ is required to be larger than
or equal to $10^{-4}$, the condition on the orbital eccentricity is
relaxed to $e \ga 0.04$.

The high-order harmonics required to reach forcing frequencies of the
order of $\sim 5.0\, \tau_{\rm dyn}^{-1}$ need very high orbital
eccentricities in order for the associated terms to contribute
non-trivially to the tide-generating potential. These high orbital
eccentricities in combination with short orbital periods yield small
periastron distances which may cause the star to overflow its
Roche-lobe when it passes through the periastron of its relative
orbit.  The critical orbital periods separating detached from
semi-detached systems were derived in Section~\ref{et} and presented
in Figs.~\ref{epsTmax_zams} and~\ref{epsTmax_msmax} for different
combinations of the binary mass ratio $q$ and the orbital eccentricity
$e$. For illustration, the orbital periods leading to semi-detached
systems in the case of the binary mass ratio $q=1$ and the orbital
eccentricity $e=0.5$ are displayed in Figs.~\ref{M1Pzams}
and~\ref{M1Pmsmax} by means of the grey-shaded background. In this
example, the critical orbital periods separating detached from
semi-detached systems increase from $\sim 1.5$ days for a $2\,M_\odot$
ZAMS star to $\sim 3$ days for a $20\,M_\odot$ ZAMS star, and from
$\sim 5$ days for a $2\,M_\odot$ TMS star to $\sim 20$ days for a
$20\,M_\odot$ TMS star. The minimum orbital eccentricity $e=0.52$
required to have a non-negligible contribution to the tide-generating
potential from the terms associated with $k$-values up to 30 can
therefore not be reached by short-period systems without becoming
semi-detached at the periastron. From Figs.~\ref{epsTmax_zams}
and~\ref{epsTmax_msmax}, it follows that the same holds true for the
other high-order harmonics required to reach forcing frequencies of
the order of $\sim 5.0\, \tau_{\rm dyn}^{-1}$. The situation improves
somewhat for binaries with mass ratios $q$ smaller than unity, but
even for $q=0.1$ the systems can barely avoid a semi-detached state
when the stars are located at the periastron of their relative orbit.

Next, we turn our attention to the forcing frequencies of the order of
$0.5\,\tau_{\rm dyn}^{-1}$ associated with the lower-order harmonics
in Expansion~(\ref{pot}) of the tide-generating potential. These
harmonics provide non-negligible contributions to the tidal potential
for low to moderate values of the orbital eccentricity, so that the
occurrence of Roche-lobe overflow at the periastron is less of an
issue.

In Section~\ref{et}, we have shown that the order of magnitude of any
given term in the expansion of the tide-generating potential is not
only determined by the orbital eccentricity $e$, but also by the
parameter $\varepsilon_T$ which is proportional to the ratio of the
tidal force to the gravity at the star's equator. Since the parameter
$\varepsilon_T$ is the dominant factor in binaries with low to
moderate orbital eccentricities, we neglect the role of the
eccentricity in what follows and, somewhat arbitrarily, set the limit
of what we consider as favourable conditions for the resonant
excitation of free oscillation modes to $\varepsilon_T \ga
10^{-3}$. In the case of a binary consisting of two $5\,M_\odot$ ZAMS
stars, this corresponds to an upper limit on the orbital period of
$\sim 5$ days. A similar upper limit for the occurrence of noticeable
resonances between dynamic tides and free oscillation modes is
inferred from the tidally induced radial-velocity variations presented
by Willems \& Aerts (2002) and the apsidal-motion calculations
performed by Smeyers \& Willems (2001).

The maximum value of $k$ for which the orbital periods represented by
the solid lines in Figs.~\ref{M1Pzams} and~\ref{M1Pmsmax} give rise to
forcing frequencies of the order of $0.5\,\tau_{\rm dyn}^{-1}$ and for
which $\varepsilon_T \ga 10^{-3}$ are obtained from Eq.~(\ref{epsT6})
as
\begin{equation}
k_{\rm max} \approx 15.8\, \left( {q \over {1+q}} \right)^{1/2}.
  \label{epsT7}
\end{equation}
The corresponding orbital periods are
\begin{equation}
P_{\rm orb, max} \approx 198.7\, \tau_{\rm dyn}
  \left( {q \over {1+q}} \right)^{1/2}.  \label{epsT8}
\end{equation}
In the case of a binary with a mass ratio $q=1$, the upper limits take
the values $k_{\rm max} \approx 11$ and $P_{\rm orb, max} \approx
140\, \tau_{\rm dyn}$. They change by less than a factor of $\sqrt{2}$
when $1/3 < q < \infty$, but decrease rapidly for binary mass ratios
smaller than $1/3$.  The minimum orbital eccentricity required for the
terms associated with $k \approx 11$ to provide a non-negligible
contribution to the tide-generating potential is $e \approx
0.25-0.4$. The longest orbital periods thus still require substantial
orbital eccentricities in order to be able to excite oscillation modes
with eigenfrequencies of the order of $0.5\,\tau_{\rm dyn}^{-1}$.

The results shown in Figs.~\ref{M1Pzams} and~\ref{M1Pmsmax} can
furthermore be extrapolated to resonances between dynamic tides and
free oscillation modes with eigenfrequencies smaller than
$0.5\,\tau_{\rm dyn}^{-1}$. From Eq.~(\ref{rng2}) it follows that the
orbital periods giving rise to such resonances increase with
decreasing order of magnitude of the eigenfrequencies, so that the
condition $\varepsilon_T \ga 10^{-3}$ becomes more and more stringent
until it can no longer be fulfilled for any value of $k$ and $q$. The
smallest multiple $f$ of the star's dynamical time scale for which the
orbital periods resulting from Eq.~(\ref{rng2}) yield values of
$\varepsilon_T$ larger than $10^{-3}$ is given by
\begin{equation}
f_{\rm min} \approx 0.03 \left( {{1+q} \over q} \right)^{1/2}
  \label{fmin}
\end{equation}
[see Eq.~(\ref{epsT6})].  For binaries with a mass ratio $q=1$, this
imposes a lower limit of $\sim 0.05\,\tau_{\rm dyn}^{-1}$ on the
frequency range favourable for resonances between dynamic tides and
free oscillation modes. In the case of a binary with a mass ratio
$q=0.1$, the lower limit increases to $\sim 0.1\,\tau_{\rm
dyn}^{-1}$. The conditions for the resonant excitation of free
oscillation modes $g^+$ thus rapidly become less favourable with
decreasing order of magnitude of the involved frequencies.  

So far, we only considered slowly rotating stars for which $\Omega \ll
\tau_{\rm dyn}^{-1}$. In more rapidly rotating stars the rotational
angular velocity $\Omega$ may significantly affect the determination of
the forcing angular frequencies induced by the companion. Resonances
with eigenfrequencies of the order of $f$ times the inverse of star's
dynamical time scale then occur for orbital periods 
\begin{equation}
P_{\rm orb} \approx {{2\, \pi\, k} \over {f - m\, \Omega\,
  \tau_{\rm dyn}}}\, \tau_{\rm dyn}.  \label{rng2b}
\end{equation}
For $m=0$, the equation reduces to Eq.~(\ref{rng2}) so that the
results presented in Figs.~\ref{M1Pzams} and~\ref{M1Pmsmax} generally
remain valid. However, since the coefficients $c_{2,0,k}$ decrease
more rapidly with increasing values of $k$ than the coefficients
$c_{2,-2,k}$ the associated terms in the expansion of the
tide-generating potential require a higher orbital eccentricity in
order to provide a non-negligible contribution. For $m=-2$, on the
other hand, the periods represented by the solid lines in
Figs.~\ref{M1Pzams} and~\ref{M1Pmsmax} decrease with increasing
rotational angular velocities $\Omega$.  Consequently, for both $m=0$
and $m=-2$, the upper limit of $5.0\,\tau_{\rm dyn}^{-1}$ on the
frequency range that is favourable for the resonant excitation of free
oscillation modes becomes even more stringent for rapidly rotating
stars than it was for more slowly rotating stars. The upper limit on
the range of orbital periods favourable for the resonant excitation of
oscillation modes with eigenfrequencies of the order of
$0.5\,\tau_{\rm dyn}^{-1}$ is not affected by the rotational angular
velocity $\Omega$, although the value of $k$ for which the upper
limit is reached increases with increasing values of $\Omega$. Our
qualitative results thus remain valid for rotation rates that are
somewhat higher than implied by Eq.~(\ref{omega}). In the case of very
fast rotators, the situation becomes significantly more complicated
since the forcing frequencies may take negative as well as positive
values. We refer to Willems \& Claret (2003) for a more detailed
account on the behaviour of the forcing frequencies in such stars.

Finally, we note that for a rotating star the frequencies and 
properties of $g^+$-modes with eigenfrequencies of the order of the
rotational angular velocity $\Omega$ may be significantly altered with
respect to those of a non-rotating star. In particular, stellar
rotation lifts the degeneracy of the eigenfrequencies with respect to
the azimuthal number $m$, so that more modes become available which
may be resonantly excited (e.g. Unno et al. 1989). Stellar rotation
furthermore induces additional low-frequency modes known as inertial
modes and $r$-modes which have frequencies of the order of the star's
rotational angular velocity $\Omega$. The resonant excitation of these
modes may lead to comparably important effects as the resonant
excitation of free oscillation modes $g^+$ (Rocca 1982, Savonije \&
Papaloizou 1997, Papaloizou \& Savonije 1997). 

\section{Concluding remarks}

In this paper, we initiated a systematic study aimed at exploring the
range of stellar and orbital parameters leading to favourable
conditions for the resonant excitation of $g^+$-modes by tides in
close binary components. We considered slowly rotating stars with
masses between 2 and $20\,M_\odot$, and evolutionary stages at the
beginning and at the end of the main sequence.

A crucial ingredient in the investigation is the expansion of the
tide-generating potential in Fourier series in terms of the mean
Keplerian motion. The expansion induces an infinite number of forcing
frequencies in the star which may be close to the eigenfrequencies of
the star's free modes of oscillation. In practice, the rapid decrease
of the Fourier coefficients limits the number of forcing frequencies
to a finite number which increases with increasing values of the
orbital eccentricity. The number can be derived from the variations of
the dominant Fourier coefficients shown in Figs.~\ref{c-2}--\ref{c2}.

For given orbital and rotational periods,  
the finite number of non-trivial terms in the expansion of the
tide-generating potential associated with a given value of the orbital
eccentricity puts an upper limit on the forcing frequencies induced in
a star, which in turn puts an upper limit on the eigenfrequencies of
the oscillation modes that may be resonantly excited. The limit is of
the order of five times the inverse of the star's dynamical time scale
and is particularly stringent for the $f$-mode and the lowest-order
$g^+$-modes in evolved low-mass main-sequence stars (see
Fig.~\ref{freqs}). It also contributes to the difficulties to excite
$p$-modes by tides in close binaries (see also Uytterhoeven et
al. 2003, in preparation).

Resonances between dynamic tides and $g^+$-modes with frequencies of
the order of half the inverse of the star's dynamical time scale on
the other hand are constrained by the rapid decrease of the strength
of the tidal force with increasing orbital separations. When the limit
separating favourable from less favourable conditions is set at
$\varepsilon_T \approx 10^{-3}$, where $\varepsilon_T$ is a
measure for the ratio of the tidal force to the gravity at the star's
equator, the highest order harmonics in the expansion of the
tide-generating potential that may give rise to such resonances are
given by Eq.~(\ref{epsT7}). This corresponds to an upper limit on the
orbital period given by Eq.~(\ref{epsT8}). For oscillation modes 
$g^+$ with eigenfrequencies much smaller than half the inverse of the
star's dynamical time scale, the conditions for resonant excitation
quickly become less favourable.
However, these very low frequencies are only reached by $g^+$-modes of
very high radial order, so that this behaviour does not impose severe
constraints on the parameter space available for the resonant
excitation of free oscillation modes.

\section*{Acknowledgements}
I express my sincere thanks to Antonio Claret for providing a
set of theoretical stellar models and to Jarrod Hurley, Onno Pols, and
Chris Tout for sharing their SSE software package. The anonymous
referee is acknowledged for useful remarks which led to an improvement
of the paper. This research was supported by the British Particle
Physics and Astronomy Research Council (PPARC) and made use of NASA's
Astrophysics Data System Bibliographic Services.

\bsp

\label{lastpage}

\end{document}